\documentclass[10pt,conference]{IEEEtran}%
\usepackage{amsmath}
\usepackage{bbm}
\usepackage{graphicx}
\usepackage{epsfig}
\usepackage{array}
\usepackage{psfrag}
\usepackage{amssymb}
\usepackage{mathdots}
\usepackage{color}
\usepackage{pstricks,pst-node,pst-text,pst-3d,pst-plot}
\usepackage{psfrag}
\usepackage{enumerate}
\usepackage{url,cite}
\usepackage{amsfonts}%
\usepackage{amsthm}
\usepackage{dsfont}%
\usepackage{verbatim}%
\usepackage{setspace}
\usepackage{float}
\usepackage{url}
\usepackage{fancyhdr}
\usepackage{bm}
\usepackage{lipsum} 
\newcommand\blfootnote[1]{%
  \begingroup
  \renewcommand\thefootnote{}\footnote{#1}%
  \addtocounter{footnote}{-1}%
\endgroup
}
\usepackage[hmargin=0.625in , bottom=1in, top=0.75in]{geometry}
\usepackage{algorithm}
\usepackage{algorithmic}
\newtheorem{thm}{Theorem}
\newtheorem{lma}{Lemma}
\newtheorem{Def}{Definition}
\DeclareMathOperator{\E}{\mathbb{E}}
\newcommand{\bP}{\overline{P}}

\newcommand{\lb}{\left (}
\newcommand{\rb}{\right )}

\newcommand{\script}[1]{{\mathcal {#1}}}

\newcommand{\Pmax}{P_{\rm max}}
\newcommand{\Pmin}{P_{\rm min}}

\newcommand{\Iavg}{I_{\rm avg}}
\newcommand{\gmax}{g_{{\rm max},i}}
\newcommand{\fgammai}{f_{\gamma_i}}
\newcommand{\fgi}{f_{g_i}}
\newcommand{\EE}[1]{\E \left[ #1 \right]}
\newcommand{\EEU}[1]{\E_{\bfU(k)} \left[ #1 \right]}
\newcommand{\EEY}[1]{\E_{\bfY(k)} \left[ #1 \right]}

\newcommand{\bgi}{\overline{g}_i}
\newcommand{\bW}{\overline{W}}

\newcommand{\Pit}{P_i^{(t)}}
\newcommand{\git}{g_i^{(t)}}
\newcommand{\bfP}{{\bf P}}
\newcommand{\bfPst}{{\bf P}^*}

\newcommand{\bfpi}{{\bm{\pi}}}
\newcommand{\bfpist}{{\bm{\pi}}^*}

\newcommand{\bgamma}{\overline{\gamma}}

\newcommand{\Xvq}{\{X(k)\}_{k=0}^\infty}
\newcommand{\Yivq}{\{Y_i(k)\}_{k=0}^\infty}

\newcommand{\bfY}{{\bf Y}}
\newcommand{\bfU}{{\bf U}}
\newcommand{\bfr}{{\bf r}}

\newcommand{\parFdef}[1]{\triangleq [#1_1(k),\cdots,#1_N(k)]^T}

\newcommand{\DOAC}{\emph{DOAC}}

\newcommand{\brho}{\overline{\rho}}
\newcommand{\Ri}{R_i^{(t)}}
\newcommand{\FDurK}{T_k}
\newcommand{\Wup}{\tilde{W}_{\pi_j}}

\newcommand{\dOne}{60\Ts}
\newcommand{\dTwo}{45\Ts}

\newcommand{\tP}{{\bf \tilde{P}}}
\newcommand{\tPsi}{\tilde{\Psi}}
\newcommand{\trho}{\tilde{\rho}}
\newcommand{\tS}{{\bf \tilde{S}}}
\newcommand{\sS}{\script{S}}

\newcommand{\brhomax}{\brho^{\rm max}}

\newcommand{\Rmax}{R_{{\rm max},i}}
\newcommand{\gammamax}{\gamma_{{\rm max},i}}

\newcommand{\Ts}{T}

\newcommand{\HOL}{H}

\newcommand{\SU}[1]{{\rm SU}_{#1}}

\begin{document}
\title{Delay Optimal Joint Scheduling-and-Power-Control for Cognitive Radio Uplinks}

\author{Ahmed Ewaisha, Cihan Tepedelenlio\u{g}lu\\
\small{School of Electrical, Computer, and Energy Engineering, Arizona State University, USA.}\\
\small{Email:\{ewaisha, cihan\}@asu.edu}\\
}
\maketitle
\blfootnote{The work in this paper has been supported by NSF Grant CCF-1117041.}
\begin{abstract}
An uplink cognitive radio system with a single primary user (PU) and multiple secondary users (SUs) is considered. The SUs have an individual average delay constraint and an aggregate average interference constraint to the PU. If the interference channels between the SUs and the PU are statistically different due to the different physical locations of the SUs, the SUs will experience different delay performances. This is because SUs located closer to the PU transmit with lower power levels. A dynamic scheduling-and-power-allocation policy that uses the dynamic programming, is proposed. The proposed policy can provide the required average delay guarantees to all SUs irrespective of their location as well as protect the PU from harmful interference. The policy is shown to be asymptotically delay optimal in the light traffic regime. Motivated, by the high complexity of the dynamic programming algorithm in the optimal policy we exploit the structure of the problem's solution to present an alternative suboptimal policy. Through simulations we show that in both light and heavy traffic regimes, the delay performance of the suboptimal policy is within $0.3\%$ from the optimal policy and both outperforming existing methods.
\end{abstract}

\section{Introduction}

The problem of scarcity in the radio spectrum has led to a wide interest in cognitive radio (CR) networks. CRs refer to devices that coexist with the licensed spectrum owners called the primary users (PUs). CRs are capable of dynamically adjusting their transmission parameters according to the environment to avoid harmful interference to the PUs.

In real-time applications, such as audio and video conference calls, one of the most effective QoS metrics is the average time a packet spends in the queue before being fully transmitted, quantified by average queuing delay. This is because packets are expected to arrive to the destination before a prespecified deadline. The average queuing delay needs to be as small as possible to prevent jitter and to guarantee acceptable QoS for these applications \cite{shakkottai2002scheduling,kang2013performance}. This delay can be controlled via efficient scheduling and power control algorithms. 


The problem of scheduling and/or power control for CR systems has been widely studied in the literature (see e.g., \cite{Letaief_PU_Known_Location,6924778,NEP_Distributed,Ewaisha_TVT2015,Iter_Bit_Allocation_OFDM,6464638,Neely_CNC_2009}, and the references therein). An uplink CR system is considered in \cite{Letaief_PU_Known_Location} where the authors propose a scheduling algorithm that minimizes the interference to the PU where all users' locations including the PU's are known to the secondary base station. The objective in \cite{6924778} is to maximize the total network's welfare. While this could give good performance in networks with users having statistically homogeneous channels, the users might experience degraded QoS when their channels are heterogeneous. In \cite{NEP_Distributed} a distributed scheduling algorithm that uses an on-off rate adaptation scheme is proposed. The authors of \cite{Ewaisha_TVT2015} propose a closed-form water-filling-like power allocation policy to maximize the CR system's per-user throughput. 
The work in \cite{Neely_CNC_2009} proposes a scheduling algorithm to maximize the capacity region subject to a collision constraint on the PUs. The algorithms proposed in all these works aim at optimizing the throughput for SUs while protecting the PUs from interference. However, providing guarantees on the queuing delay in CR systems was not the goal of these works. While \cite{li2011delay} proposes joint scheduling-and-power-allocation policy to minimize the of jobs scheduled at CPUs, power allocation in the presence of wireless channels is more challenging since users need to allocate the power based on the fading coefficient, an aspect that does not exist in CPUs.


The authors of \cite{Ewaisha_Asilomar_2015} propose a scheduling policy to minimize the sum of SUs' average delays. The work is extended in \cite{Ewaisha_Asilomar_2016} where a joint power allocation and scheduling policy was proposed to address the problem under an instantaneous interference constraint. In CRs, power control dictates adhering to PU's, instantaneous or average, interference constraints. In this paper, we extend the work in \cite{Ewaisha_Asilomar_2016} to study the problem under average interference constraints, for the first time in the literature. Specifically, we consider the joint scheduling and power control problem of minimizing the sum average delay of SUs subject to an average interference constraint at the PU. Our model assumes a general fading model as opposed to on-off channels considered in the literature. The novel contributions of this paper include: i) proposing a joint power-control and scheduling policy that is optimal with respect to the sum average delay of SUs under an average interference constraint; ii) proposing a policy using and Lyapunov analysis to show that it meets the heterogeneous per-user average delay requirements; and iii) proposing a suboptimal low complexity alternative that is shown in the simulations to be close to optimal. Through simulations, we show that conventional existing algorithms as the max-weight and the Carrier-Sense-Multiple-Access (CSMA) scheduling algorithms, if applied directly, can degrade the quality of service of both SUs as well as the PUs.

The rest of the paper is organized as follows. The system model and the problem formulation are presented in Section \ref{Model}. The proposed policy and its optimality and complexity are presented in Section \ref{Proposed_Algorithm}. Section \ref{Results} presents our extensive simulation results. The paper is concluded in Section \ref{Conclusion}.

\section{System Model and Problem Statement}
\label{Model}
\begin{figure}%
\centering
\includegraphics[width=0.65\columnwidth]{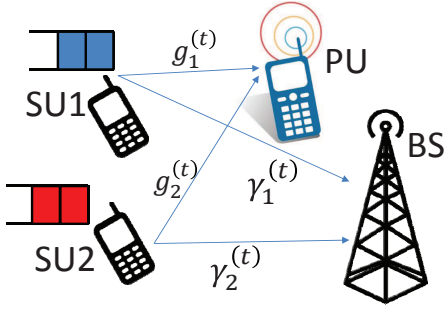}%
\caption{An uplink CR system with $N$ SUs (in this figure $N=2$) and one PU receiver. There exists an interference link between each SU and the PU that is assumed to be using the channel continuously.}%
\label{Cell_Fig}%
\end{figure}

We assume a CR system consisting of a single secondary base station (BS) serving $N$ secondary users (SUs) indexed by the set $\script{N}\triangleq \{1,\cdots N\}$ (Fig. \ref{Cell_Fig}). We are considering the uplink phase. SUs share a single frequency channel with a single PU that has licensed access to this channel. The CR system operates in an underlay fashion where the PU is using the channel continuously at all times. SUs are allowed to transmit as long as the interference received by the PU averaged over a large duration of time does not exceed a prespecified threshold $\Iavg$. Moreover, we assume that no more than one SU at a time slot should be assigned the channel.

\subsection{Channel and Interference Model}
We assume a time slotted structure where each slot is of duration $\Ts$ seconds. The channel between $\SU{i}$ and the BS (and that between $\SU{i}$ and the PU) is block fading with instantaneous power gain $\gamma_i^{(t)}$ (with gain $g_i^{(t)}$), at time slot $t$, following the probability mass function $\fgammai(\gamma)$ with mean $\bgamma_i$ ($\fgi(g)$ with mean $\bgi$) and i.i.d. across time slots, and $\gammamax$ ($\gmax$) is the maximum gain $\gamma_i^{(t)}$ ($\git$) could take. The channel gains are statistically independent and heterogeneous across SUs. We assume perfect knowledge of $\gamma_i^{(t)}$ and $g_i^{(t)}$ at the beginning of slot $t$ through some channel estimation phase that is out of the scope of this work (see \cite[Section VI]{haykin2005cognitive} and \cite{Bari13ciss,Bari13asilomar,Bari14asilomar,Bari15spl,Bari15ciss1,Bari15asilomar1} for different channel estimation techniques in CRs).
SUs use a rate adaptation scheme based on the channel gain $\gamma_i^{(t)}$. The transmission rate of $\SU{i}$ at time slot $t$ is $\Ri(\Pit)\triangleq\log \lb 1+P_i^{(t)}\gamma_i^{(t)} \rb$ bits, where $\Pmin\leq P_i^{(t)}\leq\Pmax$ is the power by which $\SU{i}$ transmits its bits at slot $t$, for some minimum and maximum values $\Pmin$ and $\Pmax$. We assume that there exists a finite maximum rate $\Rmax\triangleq\log \lb 1+\Pmax\gammamax \rb$ that $\SU{i}$ cannot exceed. Moreover, we define $\mu_i(P)=\EE{\Ri(P)}/L$ packets/slot where $L$ is the number of bits per packet with $L\gg\max_i(\Rmax)$ which is a typical case for packets with large sizes as video packets \cite[Section 3.1.6.1]{semiconductor2008long}.

\subsection{Queuing Model}
\subsubsection{Arrival Process} We assume that packets arrive to the $\SU{i}$'s buffer at the beginning of each slot. Let $\script{A}_i^{(t)}$ be the number of packets arriving to $\SU{i}$'s buffer at slot $t$. We assume $\script{A}_i^{(t)}$ is a Bernoulli process with a fixed parameter $\lambda_i$ packets per time slot and packets are buffered in infinite-sized buffers.

\subsubsection{Service Process} Packets are served according to the first-come-first-serve \emph{preemptive resume} queuing discipline. Thus at slot $t$, if $\SU{i}$ is assigned the channel, it transmits $M_i^{(t)}\triangleq\min \lb \Ri(\Pit),\HOL_i(t) \rb$ bits of the head-of-line (HOL) packet of its queue, where $\HOL_i(t)$ is the remaining number of bits of the HOL packet of $\SU{i}$ at the beginning of slot $t$ and is given by
\begin{equation*}
\HOL_i(t+1)\triangleq\left\{
\begin{array}{lll}
L\mathds{1}\lb Q_i^{(t)}+\vert \script{A}_i^{(t+1)}\vert>0\rb, & M_i^{(t)}= \HOL_i(t)\\
\HOL_i(t) - M_i^{(t)}, &\mbox{otherwise}
\end{array}
\right.
\end{equation*}
where $\mathds{1}(x)=1$ if the event $x$ occurs and $0$ otherwise while $Q_i^{(t)}$ represents the number of packets in $\SU{i}$'s queue at the beginning of slot $t$ that evolves as follows $Q_i^{(t+1)}= \lb Q_i^{(t)} + \vert \script{A}_i^{(t)}\vert - S_i^{(t)} \rb^+$
where the packet service indicator $S_i^{(t)}=1$ if $\HOL_i(t)=M_i^{(t)}$ and $0$ otherwise.

The service time $s_i$ of $\SU{i}$ is the number of time slots required to transmit one packet for $\SU{i}$, excluding the service interruptions. It can be shown that $\EE{s_i}=\EE{\Ri}/L$ time slots/packet. Hence, the service time follows a general distribution that depends on the distribution of $\gamma_i^{(t)}$. The packet arriving to $\SU{i}$ at slot $t$ has a delay $W_i^{(t)}$ which is the total amount of time, in time slots, that this packet spends in the system. The time-average delay experienced by $\SU{i}$'s packets is given by \cite{li2011delay} $\bW_i \triangleq \lim_{T \rightarrow \infty} \EE{\sum_{t=1}^T{W_i^{(t)}}}/\EE{\sum_{t=1}^T{\script{A}_i^{(t)}}}$

\subsection{Frame-Based Policy}
\label{Frame_Based_Policy}
\begin{figure}
\centering
\includegraphics[width=1\columnwidth]{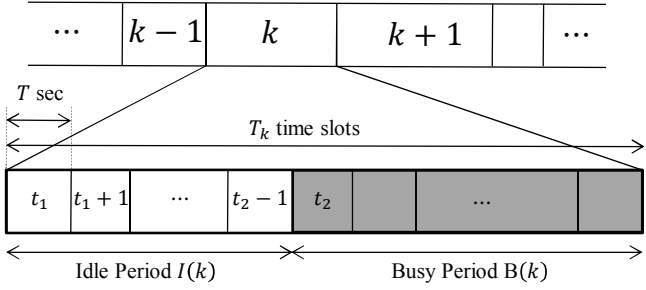}%
\caption{Frame $k$ has $\FDurK\triangleq\vert \script{F}(k)\vert$ slots with $\Ts$ seconds. Different frames can have different number of slots. Idle period is the time slots where buffers of all users are empty.}%
\label{Frame_Structure}%
\end{figure}

The idea behind our policy is to divide time into frames and update the power allocation and scheduling at the beginning of each frame. Frame $k$ consists of $T_k\triangleq\vert \script{F}(k)\vert$ consecutive time-slots, where $\script{F}(k)$ is the set containing the indices of the time slots belonging to frame $k$. Frame $k$ starts when all buffers of all users become empty in frame $k-1$ (see Fig. \ref{Frame_Structure}).

We define $\bfpi(k) \parFdef{\pi}$ where $\pi_j(k)$ is the index of the SU who is given the $j$th priority during frame $k$. Given $\bfpi(k)$, the scheduler becomes a priority scheduler with preemptive-resume priority queuing discipline \cite[pp. 205]{Bertsekas_Data_Networks}. The idea of dividing time into frames and assigning fixed priority lists for each frame was also used in \cite{li2011delay}. During frame $k$, SUs are scheduled according to some priority list $\bfpi(k)$ and each SU is assigned some power to be used when it is assigned the channel. The priority list and the power functions are fixed during the entire frame $k$ and are found at the beginning of frame $k$ based on the history of SUs' time-averaged delays and, in the case of \eqref{Prob}, the PU's suffered interference up to the end of frame $k-1$. An equivalent equation for the average delay $\bW_i$ is
\begin{equation}
\bW_i \triangleq \lim_{K \rightarrow \infty} \frac{\EE{\sum_{k=0}^K \lb\sum_{j\in \script{A}_i(k)}W_i^{(j)}\rb}}{\EE{\sum_{k=0}^{K}{\vert\script{A}_i(k)\vert}}}
\label{Delay_Frame}
\end{equation}
where $\script{A}_i(k)\triangleq\cup_{t\in\script{F}(k)}\script{A}_i^{(t)}$ is the set of all packets that arrive at SU $i$'s buffer during frame $k$.

\subsection{Problem Statement}
\label{Prob_Statement}
Each $\SU{i}$ has an average delay constraint $\bW_i \leq d_i$ that needs to be satisfied. Moreover, there is an interference constraint that the SU needs to meet in order to coexist with the PU. The main objective is to solve the following problem
\begin{equation}
\begin{array}{ll}
\underset{\{\bfpi(k)\},\{\bfP{}(k)\}}{\rm{minimize}}& \sum_{i=1}^N \bW_i \\
\label{Prob}
\rm{subject \; to} & I\triangleq\lim_{T\rightarrow \infty} \sum_{i=1}^N{\frac{1}{T}\sum_{t=1}^T{P_i^{(t)} g_i^{(t)}}}\leq \Iavg\\
& \bW_i \leq d_i\hspace{0.1in},  \hspace{0.1in} \forall i \in \script{N}\\
& \Pmin\leq P_i^{(t)} \leq\Pmax \hspace{0.05in} , \hspace{0.05in} \forall i \in \script{N} \rm{\; and \;}\forall t \geq 1,\\
& \sum_{i=1}^N{  \mathds{1} \lb P_i^{(t)}\rb} \leq 1 \hspace{0.1in}, \hspace{0.1in} \forall t \geq 1,
\end{array}
\end{equation}

Problem \eqref{Prob} is a joint power allocation and scheduling problem the solution of which guarantees the desired delay performances for all SUs. We notice that the objective function and constraints of \eqref{Prob} are expressed in terms of asymptotic time averages and cannot be solved by conventional optimization techniques. The next section proposes low complexity update policy to solve this problem and proves its optimality.

\section{Proposed Policy}
\label{Proposed_Algorithm}
We solve \eqref{Prob} by proposing online joint scheduling and power allocation policy that dynamically updates the scheduling and the power allocation. We show that this policy has a performance that comes arbitrarily close to being optimal as some control parameter $V\rightarrow\infty$.

To satisfy the delay constraints  (interference constraint) in \eqref{Prob} we set up a ``virtual queue'' associated with each delay constraint $\bW_i\leq d_i$ (interference constraint $I\leq \Iavg$). The virtual queue for $\bW_i\leq d_i$ ($I\leq \Iavg$) at frame $k$ is given by
\begin{equation}
\label{Avg_Delay_Q}
Y_i(k+1)\triangleq\lb Y_i(k)+\sum_{j\in \script{A}_i(k)}{\lb W_i^{(j)}-r_i(k)\rb} \rb^+, \text{and}
\end{equation}
\begin{equation}
\label{Avg_Interf_Q}
X(k+1)\triangleq \lb X(k) + \sum_{i=1}^N{\sum_{t\in\script{F}(k)}{P_i^{(t)} g_i^{(t)}}} -\Iavg T_k\rb^+,
\end{equation} respectively, where $r_i(k)\in[0,d_i]$ is an auxiliary variable, that is to be optimized over and $Y_i(0)\triangleq 0$, $\forall i$, while $\script{A}_i(k)$ is the set of packets arriving to $\SU{i}$ during frame $k$. We define $\bfY(k) \parFdef{Y}$. $\bfY(k+1)$ ($X(k+1)$), calculated at the end of frame $k$, represent the amount of delay (interference) exceeding $d_i$ ($\Iavg$) for $\SU{i}$ up to the beginning of frame $k+1$. We first give the following definition, then state a lemma that gives a sufficient condition on $Y_i(k)$ ($X(k)$) for the delay (interference) constraint $\bW_i\leq d_i$ ($I\leq \Iavg$) to be satisfied.
\begin{Def}
\label{Mean_Rate_Def}
A random sequence $\{Z(k)\}_{k=0}^\infty$ is said to be mean rate stable if and only if $\lim_{K\rightarrow\infty}\EE{Z(K)}/K=0$.
\end{Def}

\begin{lma}
\label{Mean_Rate_Lemma}
If $\{Y_i(k)\}_{k=0}^\infty$ ($\{X(k)\}_{k=0}^\infty$) is mean rate stable, then the constraint $\bW_i\leq d_i$ ($I\leq \Iavg$) is satisfied.
\end{lma}
\begin{proof}
Lemma 3 in \cite{li2011delay} can be modified to show that
\begin{align}
\nonumber &\frac{\EE{\sum_{k=0}^{K-1} \lb\sum_{j\in \script{A}_i(k)}W_i^{(j)}\rb}}{\EE{\sum_{k=0}^{K-1}{\vert\script{A}_i(k)\vert}}}\\
&\leq \frac{\EE{Y_i(K)}}{K}\frac{K}{\EE{\sum_{k=0}^{K-1}{\vert\script{A}_i(k)\vert}}} + \frac{\sum_{k=0}^{K-1}{\EE{\vert\script{A}_i(k)\vert {r_i(k)}}}}{\sum_{k=0}^{K-1}\EE{\vert\script{A}_i(k)\vert}}.
\label{Wait_r_i}
\end{align}
Replacing $r_i(k)$ by its upper bound $d_i$, taking the limit as $K\rightarrow\infty$ then using the mean rate stability definition and equation \eqref{Delay_Frame} completes the mean rate stability proof of $\Yivq$. Similarly we can prove the mean rate stability of $\Xvq$.
\end{proof}
Lemma \ref{Mean_Rate_Lemma} gives a condition on the virtual queue $\Yivq$ ($\Xvq$) so that the average delay (interference) constraint $\bW_i\leq d_i$ ($I\leq \Iavg$) in \eqref{Prob} is satisfied. That is, if the proposed joint power allocation and scheduling policy results in a mean rate stable $\Yivq$ ($\Xvq$), then $\bW_i\leq d_i$ ($I\leq \Iavg$).

\subsection{Optimal Policy}
\label{Algorithm_DOAC}
We first give the following useful definitions. Since the scheduling scheme in frame $k$ is a priority scheduling scheme with preemptive-resume queuing discipline, then given the priority list $\bfpi$ we can write the expected waiting time of all SUs in terms of the average residual time \cite[pp. 206]{Bertsekas_Data_Networks} $T_{\pi_j}^{\rm R}$. The waiting time of SU $\pi_j$ that is given the $j$th priority is \cite[pp. 206]{Bertsekas_Data_Networks}
\begin{equation}
\begin{array}{ll}
W_{\pi_j}&\lb P,\mu_{\pi_j}(P),\rho_{\pi_j}(P),\brho_{\pi_{j-1}},T_{\pi_j}^{\rm R}\rb\\
&\triangleq\frac{1}{\lb 1-\brho_{\pi_{j-1}}\rb}\left[\frac{1}{\mu_{\pi_j}(P)} + \frac{T_{\pi_j}^{\rm R}}{\lb 1-\brho_{\pi_{j-1}} - \rho_{\pi_j}(P)\rb}\right]
\label{Priority_Delay}
\end{array}
\end{equation}
where $\rho_i(P)\triangleq \lambda_i/\mu_i(P)$ and $\brho_{\pi_{j-1}} \triangleq \sum_{l=1}^{j-1} \rho_{\pi_l}(P_{\pi_l})$. Moreover, we define
\begin{equation}
\begin{array}{ll}
\Wup&\lb P,\rho_{\pi_j}(P),\brhomax_{\pi_{j-1}},T_{\pi_j}^{\rm R}\rb \\
&\triangleq\frac{1}{\lb 1-\brhomax_{\pi_{j-1}}\rb}\left[\frac{1}{\mu_{\pi_j}(P)} + \frac{T_{\pi_j}^{\rm R}}{\lb 1-\brhomax_{\pi_{j-1}} - \rho_{\pi_j}(P)\rb}\right]
\label{Priority_Delay_UB}
\end{array}
\end{equation}
where $\brhomax_{\pi_{j-1}}\triangleq \sum_{l=1}^{j-1} \rho_{\pi_l}\lb P_{\pi_l}^{\brhomax}\rb$, with $P_{\pi_l}^{\brhomax}\triangleq\arg \min_P \psi_{\pi_l}\lb P,\brhomax_{\pi_{l-1}}\rb$. We henceforth drop all the arguments of $\Wup$ and $W_{\pi_j}$ except $(P,\brhomax_{\pi_{j-1}})$ and $(P)$, respectively. Since, we can show that $W_{\pi_j}(P)\rightarrow\Wup(P,\brhomax_{\pi_{j-1}})$ in the light traffic regime, we will use $\Wup(P,\brhomax_{\pi_{j-1}})$ to replace $W_{\pi_j}(P)$ in our analysis. Thus the power search problem is decoupled and becomes $N$ one-dimensional searches as will be shown later.

Before presenting the {\DOAC} policy, we first discuss the idea behind it. Intuitively, a policy that solves problem \eqref{Prob} should allocate $\SU{i}$'s power and assign its priority such that $\SU{i}$'s expected delay and the expected interference to the PU is minimized. The {\DOAC} policy is defined as the policy that selects the power parameter vector $\bfP(k)\parFdef{P}$ jointly with the priority list $\bfpi(k)$ that minimizes $\Psi\triangleq\sum_{j=1}^N\psi_{\pi_j}(P_{\pi_j}(k),\brhomax_{\pi_{j-1}})$ where
\begin{equation}
\label{Optimization_Obj}
\psi_{\pi_j}(P,\brhomax_{\pi_{j-1}})\triangleq Y_{\pi_j}(k) \lambda_{\pi_j} \Wup(P,\brhomax_{\pi_{j-1}})+ X(k)\rho_{\pi_j}(P)P\bar{g}_{\pi_j}.
\end{equation}
Since the functions $\psi_{\pi_j}(P_{\pi_j}(k),\brhomax_{\pi_{j-1}})$ are decoupled, of the vector $\bfP(k)$, for all $j\in\script{N}$, we can minimize $\Psi$, for a fixed $\bfpi(k)$ vector, iteratively starting from $j=1$. This allows the use of the dynamic programming algorithm to find the optimum $\bfpi(k)$. We now present the \emph{Delay-Optimal-under-Average-Interference-Constraint} ({\DOAC}) policy and state its optimality.

{\bf {\DOAC} Policy}:
\begin{enumerate}
	\item BS executes Algorithm \ref{DOACopt} at the beginning of frame $k$.
	\item At slot $t\in\script{F}(k)$, schedule $i^{*(t)}$ who has the highest priority in the list $\bfpist(k)$.
	\item $\SU{i^{*(t)}}$ transmits $M_{i^{*(t)}}^{(t)}$ bits where $P_{i^{*(t)}}^{(t)}=P_{i^{*(t)}}^*(k)$.
	\item At end of frame $k$, BS sets $r_i(k)= d_i$ if $V<Y_i(k)\lambda_i$ and $0$ otherwise, then updates $X(k+1)$ and $Y_i(k+1)$, $\forall i\in \script{N}$.
\end{enumerate}

\begin{algorithm}
\caption{to find $\bfPst(k)$ and $\bfpist(k)$}
\begin{algorithmic}[1]
\label{DOACopt}
\STATE Define $\script{S}$ as the set of all sets formed of all subsets of $\script{N}$ and define the auxiliary functions $\tPsi(\cdot,\cdot):\script{N}\times\sS\rightarrow \mathbb{R}^+$, $\trho(\cdot):\sS\rightarrow [0,1]$, $\tS(\script{X}):\sS\rightarrow\script{N}^{\vert\script{X}\vert}$, $\tP(\script{X}):\sS\rightarrow[0,\Pmax]^{\vert\script{X}\vert}$ and $\bP(\cdot,\cdot):\sS\times\script{N}\rightarrow[0,\Pmax]$.
\STATE Initialize $\tPsi(0,\cdot)=\trho(\{\})=0$, $\tS(\{\})=\tP(\{\})=[\hspace{0.05in}]$.
\FOR{$i=1,\cdots, N$}
\STATE In stage $i$, the first $i$ priorities have been assigned to $i$ users. The corresponding priority list is denoted $[\pi_1,\cdots,\pi_i]$. In stage $i$ we have $\binom{N}{i}$ states each corresponds to a set $j$ formed from all possible combinations of $i$ elements chosen from the set $\script{N}$. We calculate $\tPsi(i,j)$ associated with each state $j$ in terms of $\tPsi(i-1,\cdot)$ obtained in stage $i-1$ as follows.
\FOR{$j\in$ all possible $i$-element sets}
\STATE At state $j\triangleq \{\pi_1,\cdots,\pi_i\}$, we have $i$ transitions, each connects it to state $j'$ in stage $i-1$, where $j'\triangleq j\backslash l$ with $l\in j$. Find the power associated with each transition $l\in j$ denoted $\bP(j,l)\triangleq\arg\min_P \psi_l(P,\trho(j\backslash l))$.
\STATE Set
\begin{align*}
& l^*=\arg\min_{l\in j}\tPsi\lb i-1,j\backslash l\rb+ \psi_l\lb\bP(j,l),\trho(j\backslash l)\rb,\\
&\tPsi(i,j)=\tPsi(i-1,j\backslash l^*)+ \psi_{l^*}\lb\bP(j,l^*),\trho(j\backslash l^*)\rb,\\
&\trho(j)=\trho\lb j\backslash l^*\rb+\rho\lb\bP(j,l^*)\rb,\\
&\tS(j)=\left [\tS\lb j\backslash l^*\rb,l^*\right]^T,\\
&\tP(j)=\left [\tP\lb j\backslash l^*\rb,\bP(j,l^*)\right]^T.
\end{align*}
\ENDFOR
\ENDFOR
\STATE Set $\bfpist(k)= \tS\lb\script{N}\rb$ and $\bfPst(k)=\tP\lb\script{N}\rb$.
\end{algorithmic}
\end{algorithm}
The {\DOAC} policy calculates $\bfpist(k)$ and $\bfPst(k)$ once at the beginning of frame $k$ then uses these two vectors throughout the frame duration. To find $\bfpist(k)$ and $\bfPst(k)$ we solve $\min_{\bfpi(k),\bfP{}(k)} \Psi$ using the dynamic programming in Algorithm \ref{DOACopt}. Its overall complexity is of $O(MN2^N)$ where $M$ is the number of iterations in a one-dimensional search. Compared to the exhaustive search this is a large complexity reduction although still high if $N$ was large. In Section \ref{Suboptimal} we propose a sub-optimal policy with low complexity.

\subsection{Motivation of {\DOAC}}
We define $\bfU(k)\triangleq [X(k) , \bfY(k)]^T$, the Lyapunov function as $L(k) \triangleq \frac{1}{2}X^2(k)+\frac{1}{2}\sum_{i=1}^N Y_i^2(k)$ and Lyapunov drift to be
\begin{equation}
\Delta (k) \triangleq \EEU{L(k+1) - L(k)}.
\label{Drift_Def_Avg}
\end{equation}
Squaring equation \eqref{Avg_Delay_Q} and \eqref{Avg_Interf_Q} then taking the conditional expectation we can get the bounds
\begin{equation}
\begin{array}{ll}
&\frac{1}{2}\E_{\bfY(k)} \left[ Y_i^2(k+1)-Y_i^2(k)\right] \leq C_{Y_i}+\\
&\hspace{0.2in}Y_i(k) \EEY{\FDurK}\lambda_i \lb \EEY{W_i^{(j)}}-r_i(k)\rb, \rm{and }
\label{Delay_Q_Sq2}
\end{array}
\end{equation}
\begin{equation}
\begin{array}{ll}
&\frac{1}{2}\E_{\bfU(k)} \left[X^2(k+1)-X^2(k)\right]\leq C_X+\\
&\hspace{0.2in}X(k)\lb\EEU{\sum_{t\in\script{F}(k)}\Pit \git}-\Iavg\EEU{\FDurK}\rb,
\label{Interf_Q_Sq1}
\end{array}
\end{equation}
where we use the bounds $\EEY{\lb \sum_{j\in \script{A}_i(k)} W_i^{(j)}\rb^2}+\EEY{\lb\sum_{j\in \script{A}_(k)}r_i(k)\rb^2}<C_{Y_i}$ and $\EEU{\lb\sum_{i=1}^N\sum_{t\in\script{F}(k)}\Pit \git\rb^2+\lb\Iavg \FDurK\rb^2}<C_X$ and omit their derivations (see \cite{Journal_Submitted} for more details). Given some fixed control parameter $V>0$, we add the penalty term $V\sum_i \EEU{r_i(k)\FDurK}$ to both sides of \eqref{Drift_Def_Avg}. Using the bounds in \eqref{Delay_Q_Sq2} and \eqref{Interf_Q_Sq1}, the drift-plus-penalty term becomes bounded by
\begin{equation}
\Delta \lb \bfU(k)\rb + V\sum_{i=1}^N \EEU{r_i(k)\FDurK}\leq C+\EEU{\FDurK}\chi(k),
\label{Drift_Plus_Penalty_Avg}
\end{equation}
where
\begin{equation}
\chi(k)\triangleq \sum_{j=1}^N \left[\lb V -Y_j(k) \lambda_j\rb r_j(k)+\psi_{\pi_j}(P_{\pi_j}(k),\brhomax_{\pi_{j-1}})\right],
\label{chi}
\end{equation}
with $\psi_{\pi_j}(P_{\pi_j}(k),\brhomax_{\pi_{j-1}})$ defined in \eqref{Optimization_Obj}. We define the {\DOAC} policy to be the policy that jointly finds $\bfr(k)$, $\bfP(k)$ and $\bfpi(k)$ that minimize $\chi(k)$ subject to the maximum power and the single-SU-per-time-slot constraints in problem \eqref{Prob}. The updates of $r_i(k)$ in Step 4 of the {\DOAC} policy minimize the first summation of $\chi(k)$. Using the fact that $\EEU{W_{\pi_l}^{(j)}}=W_{\pi_l}^{\rm up}(P_{\pi_l}(k))$, in the light traffic regime, we get $\sum_{l=1}^N\phi_{\pi_l}$. For $\{\bfP(k)\}$ and $\bfpi(k)$, we can see that $\sum_{l=1}^N\phi_{\pi_l}$ is the only term in the right side of equation \eqref{chi} that is a function of the power allocation policy $\{\bfP(k)\}$, $\forall t\in\script{F}(k)$. Consequently, $\bfPst(k)$ and $\bfpist(k)$, the optimum values for $\bfP(k)$ and $\bfpi(k)$ respectively, are ones that minimize $\sum_{l=1}^N\phi_{\pi_l}$ as given by Algorithm \ref{DOACopt}.



\begin{thm}
\label{Optimality_Avg}
If \eqref{Prob} is strictly feasible, then for any $V>0$ there exists a constant $C<\infty$ such that, in the light traffic regime, the performance of the {\DOAC} policy satisfies
\begin{equation}
\sum_{i=1}^N{\bW_i} \leq \frac{C}{V} + \sum_{i=1}^N{\bW_i^*}
\label{Optimality_Eq}
\end{equation}
where $\bW_i^*$ is $\SU{i}$'s the optimum delay when solving \eqref{Prob}. Moreover, the virtual queues $\Xvq$ and $\Yivq$ are mean rate stable $\forall i \in \script{N}$.
\end{thm}

\begin{proof}
When evaluating by the optimum policy that solves \eqref{Prob} and by the genie-aided values of $r_i(k)=W_i^*$ in the right-hand-sides (r.h.s.) of \eqref{Delay_Q_Sq2}, \eqref{Interf_Q_Sq1} and \eqref{chi}, the second line of both \eqref{Delay_Q_Sq2} and \eqref{Interf_Q_Sq1} become negative, since the optimum policy satisfies the delay and interference constraints, respectively, while \eqref{chi} becomes $\chi^{\rm opt}\triangleq V\sum_{i=1}^N\bW_i^*$. Replacing $\chi(k)$ with $\chi^{\rm opt}$ in the r.h.s. of \eqref{Drift_Plus_Penalty_Avg} we get the bound $\Delta \lb \bfU(k)\rb + V\sum_{i=1}^N \EEU{r_i(k)\FDurK}\leq C+\EEU{\FDurK}V\sum_{i=1}^N\bW_i^*$. Taking $\EE{\cdot}$ over this inequality, summing over $k=0,\cdots,K-1$, denoting $X(0)\triangleq \bfY_i(0)\triangleq 0$ for all $i\in\script{N}$, and dividing by $V\sum_{k=0}^{K-1} \EE{\FDurK}$ we get
\begin{align}
\nonumber&\frac{\EE{X^2(K)}}{\sum_{k=0}^{K-1}\EE{\FDurK}}+\sum_{i=1}^N \frac{\EE{Y_i^2(K)}}{\sum_{k=0}^{K-1} \EE{\FDurK}}+ \sum_{i=1}^N \frac{\sum_{k=0}^{K-1}\EE{r_i(k)\FDurK}}{\sum_{k=0}^{K-1}\EE{\FDurK}}\\
 &\hspace{0.2in}\overset{(a)}{\leq} \frac{aC}{V} + \sum_{i=1}^N \bW_i^*\triangleq C_1.
\label{Optimal_Eq_Avg}
\end{align}
where in the r.h.s. of inequality (a) we used $\EE{\FDurK}\geq \EE{I(k)}=1/a$, and $C_1$ is some constant that is not a function in $K$. To prove the mean rate stability of the sequence $\{Y_i(k)\}_{k=0}^\infty$ for any $i\in\script{N}$, we remove the first and third terms in the left-side of \eqref{Optimal_Eq_Avg} as well as the summation operator from the second term to obtain $\EE{Y_i^2(K)}/K \leq C_1$ $\forall i\in\script{N}$. Using Jensen's inequality we note that
\begin{equation}
\frac{\EE{Y_i(K)}}{K} \leq \sqrt{\frac{\EE{Y_i^2(K)}}{K^2}} \leq \sqrt{\frac{C_1}{K}}.
\label{Jensens}
\end{equation}
Finally, taking the limit when $K\rightarrow \infty$ completes the mean rate stability proof of $\Yivq$. Similarly we can proof the mean rate stability of $\Xvq$. On the other hand, to prove the upper bound in Theorem \ref{Optimality_Avg}, we use the fact that $r_i(k)$ and $\vert \script{A}_i(k) \vert$ are independent random variables (see step 4 in the {\DOAC}) to replace $\EE{\vert \script{A}_i(k) \vert {r_i(k)}}$ by $\lambda_i\EE{\FDurK r_i(k)}$ in equation \eqref{Wait_r_i}, then we take the limit of \eqref{Wait_r_i} as $K\rightarrow \infty$, use the mean rate stability theorem and sum over $i\in\script{N}$ to get
\begin{equation}
\begin{array}{ll}
&\sum_{i=1}^N \frac{\EE{\sum_{k=0}^{K-1} \lb\sum_{j\in \script{A}_i(k)}W_i^{(j)}\rb}}{\EE{\sum_{k=0}^{K-1}{\vert\script{A}_i(k)\vert}}}\\
&\hspace{0.2in}\leq \sum_{i=1}^N \frac{\sum_{k=0}^{K-1}\EE{r_i(k)\FDurK}}{\sum_{k=0}^{K-1}\EE{\FDurK}}\overset{(b)}{\leq} \frac{aC_Y}{V} + \sum_{i=1}^N \bW_i^*,
\label{Optimality_Eq2}
\end{array}
\end{equation}
where inequality (b) comes from removing the first summation in the left-side of \eqref{Optimal_Eq_Avg}. Taking the limit when $K\rightarrow \infty$ and using equation \eqref{Delay_Frame} completes the proof.
\end{proof}

Theorem \ref{Optimality_Avg} says that the {\DOAC} policy is optimal as $V\rightarrow\infty$, and the interference and delay constraints of \eqref{Prob} are satisfied since $\Xvq$ and $\Yivq$ are mean rate stable.

\subsection{Near-Optimal Low Complexity Algorithm}
\label{Suboptimal}
The suboptimal policy we present here depends on decoupling the search over $\bfP(k)$ and $\bfpi(k)$. Define $\Pmin\triangleq \min\{P:\sum_{i=1}^N\rho_i(P)<1\}$. Intuitively, if, for some $\pi_j\in\script{N}$, $X(k)\gg Y_{\pi_j}(k)$ then $P_{\pi_j}^*(k)$ is expected to be close to $\Pmin$ since the second term of $\psi_{\pi_j}(P,\brhomax_{\pi_{j-1}})$ dominates over the first term. Otherwise $P_{\pi_j}^*(k)\approx \Pmax$. We propose the following two-step scheduling and power allocation algorithm: 1) for each $\SU{i}$ set $P_i(k)=\Pmin$ if $X(k)>Y_{\pi_j}(k)$ and $P_i(k)=\Pmax$ otherwise; then 2) assign priorities to SUs in a descending order of $Y_i(k)\mu_i(P_i(k))$ (the $c\mu$ rule). The complexity of this algorithm is that of sorting $N$ numbers, namely $O(N\log(N))$. Simulations will show that this causes little degradation to the average delay.

\section{Simulation Results}
\label{Results}
\begin{figure}%
\centering
\includegraphics[width=1\columnwidth]{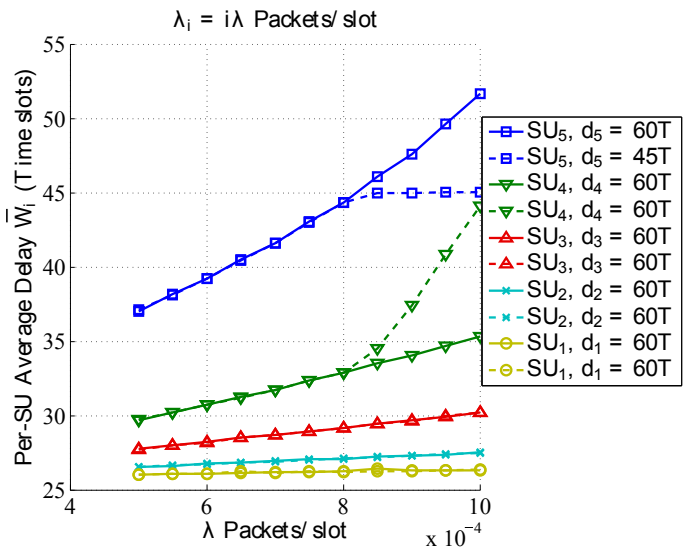}%
\caption{Average per-SU delay for both the active and inactive delay-constraint cases. Both cases are simulated using the {\DOAC} policy. $\SU{5}$ is the user with the worst channel statistics and the largest arrival rate. The {\DOAC} can guarantee a bound on $\bW_5$.}%
\label{PerUser_Delay_Avg_N5}%
\end{figure}

We simulated a system of $N=5$ SUs (Table \ref{Parameters} lists all parameter values). $\SU{i}$'s arrival rate is set to $\lambda_i=i\lambda$ for some fixed parameter $\lambda$. All SUs are having homogeneous channel conditions except $\SU{5}$ who has the highest average interference channel gain. Thus $\SU{5}$ is statistically the worst case user. We assume that SUs' delay constraints are $d_i=\dOne$ $\forall i\leq 4$, and $d_5=\dTwo$. Fig. \ref{PerUser_Delay_Avg_N5} plots the per-user delay $\bW_i$ using the {\DOAC} policy for two cases; the first is with $d_5=\dTwo$ while the second is with $d_5=\dOne$, to show the effect of an active versus an inactive delay constraint. In the active constrained case, the {\DOAC} policy guarantees that $\bW_5\leq d_5$. We conclude that the delay constraints in problem \eqref{Prob} can force SUs' delays to take any strictly feasible value.
\begin{table}
	\centering
		\caption{Simulation Parameter Values}
		\label{Parameters}
		\begin{tabular}{|c|c||c|c|}
			\cline{1-4}
			Parameter & Value & Parameter & Value \\
			\cline{1-4}
			$L$ & $1000$ bits/packet &  $\Pmax$ & 100\\
			$\fgammai(\gamma)$ & $\exp{\lb-\gamma/\bgamma_i\rb}/\bgamma_i$ & $\bgamma_i$, $\forall i$ & $1$\\
			$\fgi(g)$ & $\exp{\lb-g/\overline{g}_i\rb}/\overline{g}_i$ &$\overline{g}_i$, $i=1,2,3,4$ & $0.1$ \\
			$V$ & $100$ & $\overline{g}_5$ & $0.4$
			\\ \cline{1-4}
			\end{tabular}
\end{table}

\begin{figure}%
\centering
\includegraphics[width=0.9\columnwidth]{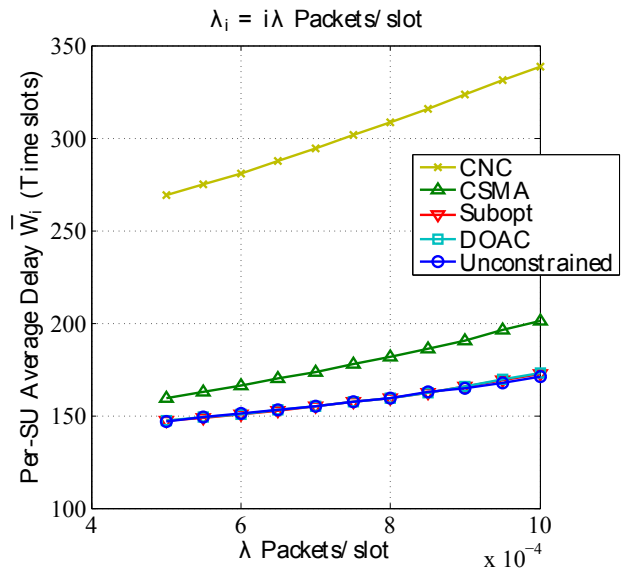}%
\caption{{\DOAC} compared to the CSMA, CNC \cite{Neely_CNC_2009} and the suboptimal algorithm.}%
\label{Sum_Delay_Avg_CSMA_CNC_Subopt_DOAC}%
\end{figure}

In Fig. \ref{Sum_Delay_Avg_CSMA_CNC_Subopt_DOAC} we compare the delay of the {\DOAC} and suboptimal policies to the Carrier-Sense-Multiple-Access (CSMA) policy and the Cognitive Network Control (CNC) policy proposed in \cite{Neely_CNC_2009}. The CSMA assigns the channel equally likely to all users while allocating the power as the {\DOAC} using a genie-aided knowledge. The CNC is a version of the MaxWeight policy. The {\DOAC} and the suboptimal policies outperform the CSMA and the CNC. This is because the proposed policies prioritize the users based on their delay and interference realizations. On the other hand, the CSMA allocates the channel to guarantee fairness of allocation across time and the CNC's goal is to maximize the achievable rate region \cite{li2011delay}. Moreover, the suboptimal policy is within $0.3\%$ of the {\DOAC} policy.

\section{Conclusion}
\label{Conclusion}
We have studied the joint scheduling and power allocation problem of an uplink multi SU CR system. We formulated the problem as a delay minimization problem in the presence of an average interference constraint to the PU, and an average delay constraint for each SU. Most of the existing literature that study this problem either assume on-off fading channels or do not provide delay guarantees to SUs. We proposed a dynamic policy that schedules SUs by dynamically updating a priority list based on the channel statistics as well as the history of the arrivals, departures and channel fading realizations. The proposed policy, referred to as the {\DOAC} policy, updates the priority list and power allocation through a dynamic programming on a per-frame basis where a single frame consists of multiple slots. We showed, through the Lyapunov optimization, that the {\DOAC} policy is asymptotically delay optimal. That is, it minimizes the sum of average delays of SUs as well as satisfying the average interference and delay constraints.

Motivated by the exponential complexity of the dynamic programming, we proposed an alternative suboptimal policy with complexity $O(N \log(N))$. Through simulations we compared this policy to the {\DOAC}, the CSMA and the CNC \cite{Neely_CNC_2009} which is a version of the MaxWeight Algorithm. Simulations show that the difference in rate between the suboptimal and the {\DOAC} policies is not more than $0.3\%$. Moreover, both policies outperform the CSMA and the CNC.



\bibliographystyle{IEEEbib}
\bibliography{MyLib}

\end{document}